\pdfoutput=1
\documentclass[11pt]{article}

\usepackage[preprint]{acl}

\usepackage{times}
\usepackage{latexsym}
\usepackage{graphicx}
\usepackage{float} 
\usepackage{booktabs}
\usepackage{makecell}
\usepackage{enumitem}
\usepackage{amsmath}
\usepackage{amssymb}
\usepackage{multirow}

\usepackage{subfigure} 
\usepackage[T1]{fontenc}
\usepackage[utf8]{inputenc}

\usepackage{microtype}

\usepackage{inconsolata}

\title{Effectiveness of ChatGPT in explaining complex medical reports to patients}

\author{\textbf{Mengxuan Sun}, \textbf{Ehud Reiter}, \textbf{Anne E Kiltie}, \\   \textbf{George Ramsay}, \textbf{Peter Murchie},   \textbf{Lisa Duncan},  \textbf{Rosalind Adam} \\
    Univeristy of Aberdeen\\
\texttt{m.sun.22@abdn.ac.uk, e.reiter@abdn.ac.uk, anne.kiltie@abdn.ac.uk,} \\
\texttt{george.ramsay@abdn.ac.uk, p.murchie@abdn.ac.uk, }\\ \texttt{lisa.duncan1@abdn.ac.uk, rosalindadam@abdn.ac.uk}
}
\usepackage{tabularx}

\begin{document}
\maketitle
\begin{abstract} 

Electronic health records contain detailed information about the medical condition of patients, but they are difficult for patients to understand even if they have access to them.  We explore whether ChatGPT (GPT 4) can help explain multidisciplinary team (MDT) reports to colorectal and prostate cancer patients. These reports are written in dense medical language and assume clinical knowledge, so they are a good test of ChatGPT’s ability to explain complex medical reports to patients.  We asked clinicians and lay people (not patients) to review ChatGPT’s explanations and responses; we also ran three focus groups (including cancer patients, caregivers, computer scientists, and clinicians) to discuss ChatGPT’s output. Our studies highlighted issues with inaccurate information, inappropriate language, limited personalization, AI distrust, and challenges integrating large language models (LLMs) into clinical workflow. These issues will need to be resolved before LLMs can be used to explain complex personal medical information to patients.

\end{abstract}

\section{Introduction}
There is considerable potential to use advanced generative LLMs, such as ChatGPT \citep{chatGPT}, in healthcare, including educating and supporting patients \citep{chiesa2024exploring,garg2023exploring}.  However, LLMs may generate incorrect information and otherwise confuse or mislead patients, which is not acceptable.

\begin{figure}[H]
\centering 
\includegraphics[width=0.5\textwidth]{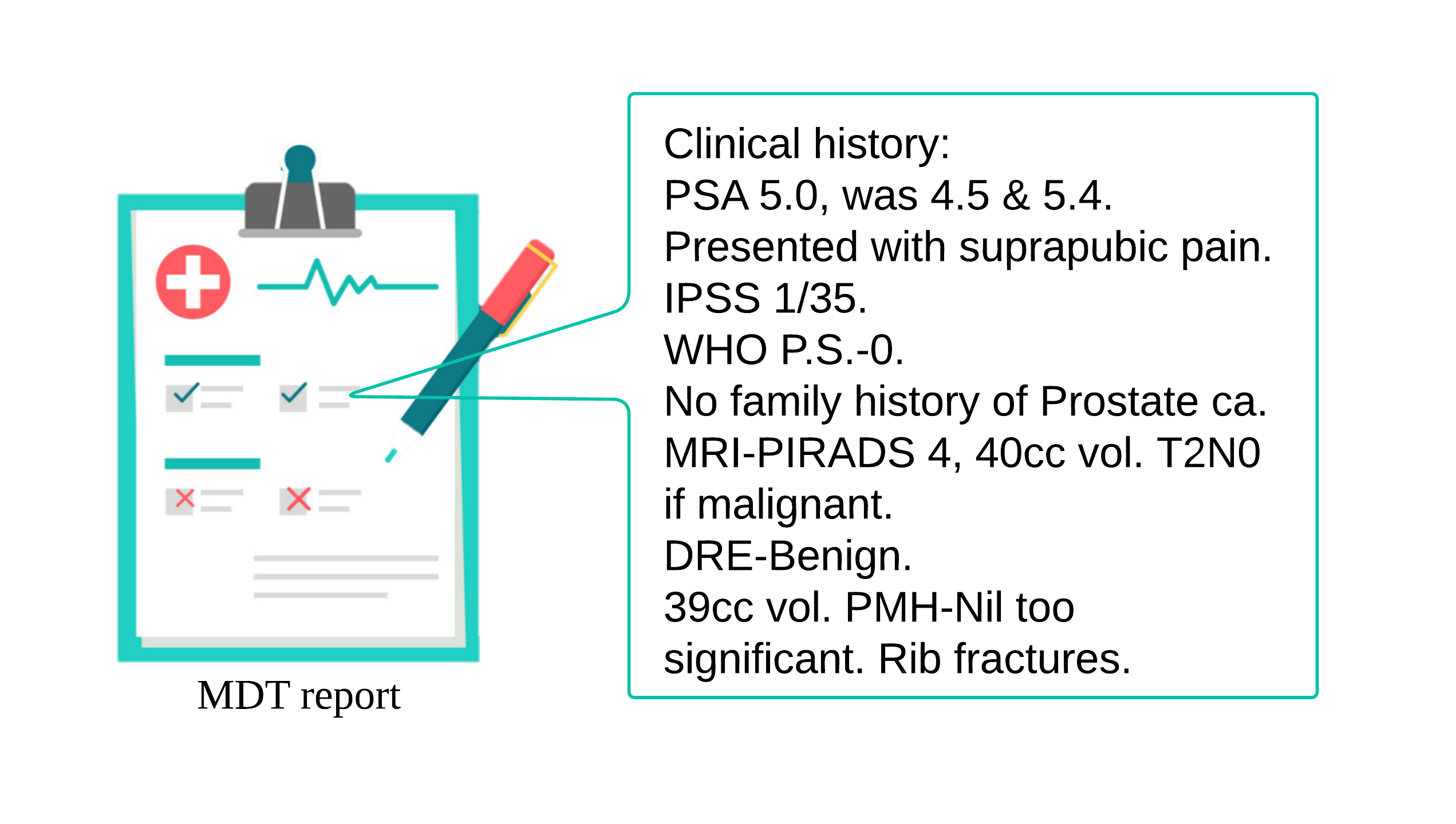} 
\caption{A fragment of clinical history from a mock MDT report. MDT reports record discussions held about patients by the extended clinical team looking after them.} 
\label{Figmdt} 
\end{figure}

Our research evaluates the ability of ChatGPT4.0 to explain complex medical reports to patients, using cancer multidisciplinary team (MDT) reports as an example (Figure \ref{Figmdt}). We asked a colorectal surgeon and clinical oncologist to create six mock MDT reports which accurately mimic reports generated for real prostate and colorectal cancer patients. We then gave each MDT report to ChatGPT and prompted it to respond to questions about the MDT in the four scenarios described in Table \ref{tab_scenarios}. These responses were analysed by the MDT reports creators, other clinicians, and lay people; they were also discussed in focus groups that included cancer patients, caregivers, computer scientists, and clinicians.

\begin{table*}[!htbp]
\centering
\begin{tabularx}{\textwidth}{lXp{7cm}p{4cm}}
\toprule
& \textbf{Scenarios} & \textbf{Description} & \textbf{Examples}\\
\hline
1 & \textbf{Patient-Explain} & The patient has the MDT report and asks ChatGPT to explain it, including the overall content and specific terms. & What does CNP mean in my report?\\
2 & \textbf{Patient-Suggest}  & The patient asks for guidance based on the information in the MDT report, including advice on lifestyle, dealing with anxiety, financial issues, etc. & Could you recommend some places where I can find support groups for people with the same type of cancer? I live in xxx.\\
3 & \textbf{Doctor-Explain} & The clinician asks ChatGPT to draft an email (which the clinician can edit) to explain the MDT results to a patient. & Write an email to this patient to inform him of MDT result \\
4 & \textbf{Doctor-Suggest} & The clinician asks ChatGPT for recommendations to help patients and outline treatment plans. & Outline a treatment plan for this patient\\
\bottomrule
\end{tabularx}
\caption{Experimental scenarios for patients and clinicians for queries to ChatGPT about complex medical reports.}
\label{tab_scenarios}
\end{table*}

We used these results to investigate two research questions, the first one is our main focus :

\textbf{Question 1}: What are the challenges of using ChatGPT to explain and provide complex medical-related information to patients?

\textbf{Question 2}: How can we address these challenges to enhance LLM's effectiveness in assisting patients with medical report explanations?

\section{Related work}

\subsection{The role of ChatGPT in handling complex cancer reports}
The advanced natural language capabilities of ChatGPT hold substantial promise in processing complex medical information, such as enhancing the efficiency of doctor-patient communication and optimizing treatment planning. \citet{Choo:23} conducted a preliminary analysis of ChatGPT's ability to formulate a management plan following oncological principles for patients with colorectal cancer, finding that the recommendation concordance rate between ChatGPT and the MDT was 86.7\%. 

%\citet{derton2023natural} analysed the linguistic features among different demographic groups and applied a variational autoencoder topic model to explore social needs in cancer patients’ notes. They found that different groups hold different typical topics.
Several studies have investigated the use of ChatGPT to assist doctors in multidisciplinary tumour board decision-making \cite{Lukac:23, Ulloa:23, Haemmerli:23, Sorin:2023}. They all reported positive results in some specific question categories but were not equivalent to the expertise of actual medical professionals.

Studies found that extracting information from cancer radiology reports can provide valuable assistance to physicians \cite{fink:23,hu2024zero} and ChatGPT has the potential to generate easy-to-understand radiology reports \citep{chung2023feasibility,jeblick2023chatgpt}.  \citet{nakamura2023chatgpt} tested ChatGPT for automating lung cancer staging based on radiology reports (accuracy 52.2\%, 78.9\%, and 86.7\% for the T, N, and M categories), indicating it’s potential to become a valuable tool.

\begin{table*}[!htbp]
\centering
\begin{tabularx}{\textwidth}{lXXp{2.5cm}X}
\toprule
& \textbf{Prompts example} & \textbf{Source} & \textbf{Scenario} \\
\hline
1 & What is a TRUS Biopsy? Why do I need to do? & Questions to Ask Your Health Care Team \cite{cancernet} & Patient-Explain\\
2 & I was diagnosed with colorectal cancer, and they had an MDT meeting about me on Feb 28th, and I’ve heard nothing. I'm just looking to see if this is normal.  & Bowel (colon and rectal) cancer forum \cite{macmsupportonlinef} & Patient-Suggest\\
3 & Write an email for the patient to inform his MDT results. & Authors' hypothesis from the patients perspective & Doctor-Explain\\
4 & What is the patient's prognosis and what should he prepare for the following treatment? & Doctor's expertise & Doctor-suggest \\
\bottomrule
\hline
\end{tabularx}
\caption{Example prompts from four sources with corresponding scenarios}
\label{tab1}
\end{table*}

\subsection{Evaluation of LLMs in medicine}
Evaluating LLMs in medicine has become a critical area of research, aiming to assess their accuracy and reliability in delivering medical information and supporting clinical practices. A few studies have reported that ChatGPT can pass medical examinations, which can assist in medical education \cite{kung2023performance, gilson2023does, kasai2023evaluating, madrid2023harnessing}. \citet{koopman2023dr} also used a public health dataset to evaluate different prompts' impact on health answer correctness. The accuracy of ChatGPT was 80\% but reduced to 63\% with additional evidence. 

Two studies mainly relied on feedback from human experts for evaluation. \citet{mehnen2023chatgpt} evaluated the diagnostic accuracy of ChatGPT with medical experts on 50 clinical case vignettes. The results show that only 40\% cases were solved with the first suggestion. \citet{tang2023evaluating} found that LLMs may produce factually inconsistent summaries and make overly convincing or overly assertive statements, potentially causing misinformation-related harm. Furthermore, \citet{balloccu2024leak} investigated data contamination problems of ChatGPT and reported that 263 distinct datasets have been exposed to the models that may highly improve the evaluation results.

\citet{tu2024towards} designed a framework AMIE for evaluating clinically meaningful axes of performance, including history-taking, diagnostic accuracy, management reasoning, communication skills and empathy. AMIE utilizes a unique self-play-based simulated environment with automated feedback mechanisms for scaling learning across various disease conditions, specialties and contexts. Both specialist physicians(28/32) and patients(24/26) reported positive performance on axes.

Three quantitative studies were conducted to evaluate the ability of LLMs to respond to cancer questions using private data.  \citet{chen2023use} illustrated that GPT4 drafts were overall helpful and safe for cancer patients’ questions, and human responses were more likely to recommend direct clinical action than LLMs. Experiments \citep{chen2024effect} also show that LLMs can reduce physician workload, improve response consistency across physician responses, and enhance the informativeness and educational value of responses towards patients. \citet{tariq2024domain} developed an oncology-specific LLM and evaluated the clinical concept coverage, clinical information retrieval recall, correctness, completeness, and relevance of the prostate question-answering task. The domain-specific model showed high relevance (91\%) but insufficient completeness (average 40\%). 

Quantitative benchmark studies provide valuable references, highlighting the performance of LLMs in comparison to gold standards on certain tasks. However, the complexity of clinical issues cannot solely be determined by binary outcomes and percentages. Qualitative analysis is required to capture patient opinions including potential risks.

\section{Methods}

\begin{figure*}
\centering 
\includegraphics[width=1\textwidth]{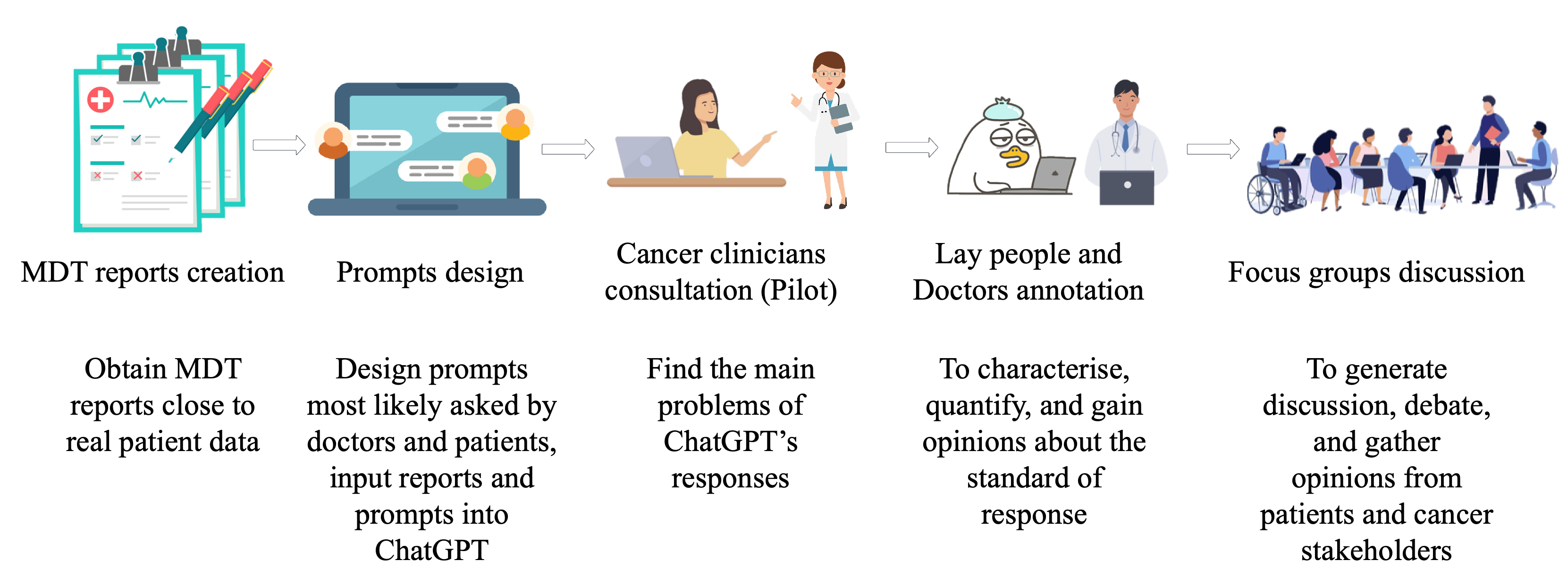} 
\caption{Methods for the study. The description below each picture is the purpose} 
\label{Figflowchart} 
\end{figure*}

\subsection{MDT reports}
A multidisciplinary meeting (MDT) is a meeting of a group of medical practitioners from different disciplinary backgrounds (e.g., oncologist, surgeon, specialist nurse, radiologist, pathologist) that work together to make decisions about patients based on clinical results and evaluations  \citep{NHSmdt}. MDT meetings are recommended as the most effective way to reach treatment decisions and are the gold standard for cancer care management\citep{taberna2020multidisciplinary}. An MDT report is the summary of the MDT meeting. The MDT report contains the patient’s information, clinical history, investigation results (e.g. radiology reports, biopsy/pathology results), and the management plan. The content is written in medical language and often contains medical jargon that requires medical expertise to interpret. Fig\ref{Figmdt} shows an excerpt from a mock MDT report as an example. 
Because of the ethical challenges in using real MDTs, we asked two senior cancer clinicians to create six fictitious MDT reports based on their expertise, ensuring that the mock reports closely resembled those of real patients but without containing any patient-identifiable data. There were four colorectal cancer reports and two prostate cancer reports, involving patients of different genders and ages, including complex cases of recurrence after treatment.

\subsection{Prompt design and test}
Multiple sources informed our prompt design: 
\begin{itemize}[itemsep=0.5em, topsep=0.5em]
\item {Frequently asked questions (FAQs) from patient resources on widely utilized cancer websites such as Cancer.net\cite{cancernet}, the National Cancer Institute \cite{NCI}, and Macmillan \cite{macmsupport}}
\item {Cancer online forums including Prostate Cancer UK \cite{prostuk}, Macmillan Cancer Support online forum \cite{macmsupportonlinef}, and Cancer Research UK \cite{cancerreasechuk}}
\item {Clinicians' experiences with common cancer patient question.}
\item{Potential patient questions envisioned by the authors.}
\end{itemize}
Table~\ref{tab1} presents example prompts from these sources corresponding to each scenario. All prompts were tailored to the MDT report of the present patient, with a clinician reviewing the language and practicality. 
The experiment tested the web version of ChatGPT-4. The pilot experiments were conducted from October to December 2023 and the content presented to doctors and patients was based on the version created from January to February 2024.

\subsection{Evaluation method}

Three evaluation methods were used in this study:
\begin{itemize}[itemsep=0.3em, topsep=0.3em]
\item {Pilot: the cancer clinicians who created the mock MDT reports checked ChatGPT responses for problems.}
\item {Clinicians and lay people (not cancer patients) commented on annotated the ChatGPT responses and also filled out a short questionnaire.}
\item {Responses were discussed in three focus groups of a co-design workshop, which included cancer patients, caregivers, computer scientists and clinicians. Figure \ref{Figflowchart} shows our workflow.}
\end{itemize}

\textbf{1) Pilot:} The oncologists who wrote the mock MDTs examined ChatGPT responses to 184 prompts (questions) about the MDTs; Appendix~\ref {sec:pilot} shows examples.  The goal was to identify and characterise errors and other problems and develop categories for types of issues.

\textbf{2) Annotation:} 
We created questionnaires based on the problems identified during the pilot study. Each questionnaire was based on one of the scenarios described in Table 1.  Subjects were shown the MDT reports, the questions asked to ChatGPT and ChatGPT’s responses, and were asked to respond to mark confusing expressions, provide reasons and give opinions on how to be more acceptable to patients.  An example is given in Appendix ~\ref {sec:MDTQ}.  They were also asked to respond to the following questions using a 5-point Likert scale (from strongly disagree to strongly agree):
\begin{itemize}[itemsep=0.3em, topsep=0.3em]
\item {I think ChatGPT handled the question well. (all subjects)}
\item {I would accept it if my doctor answered like above. (for lay subjects only)}
\item {I think the output from ChatGPT meets the standard required for implementation within clinical care. (for clinician subjects only)}
\end{itemize}
We involved five clinicians (two cancer clinicians and three general practitioners) and eight laypeople for this study. The lay people were not themselves cancer patients and were asked to assume the identity of a person with colorectal or prostate cancer.

\setlength{\tabcolsep}{5pt}
\renewcommand{\arraystretch}{1}
\setlength{\tabcolsep}{2pt}
\begin{table*}[ht]
\centering
\begin{tabular}{ccccccc}
%\begin{tabular}{c|c|c|c|c|c|c}
\toprule
%&& \multicolumn{4}{|c|}{\textbf{Scenarios}}&  \\
%\hline
&&\textbf{Patient-Explain} & \textbf{Patient-Suggest} & \textbf{Doctor-Explain} &\textbf{Doctor-Suggest} &\textbf{Overall}\\ 
\hline
\multirow{2}{*}{\textbf{Statistics(n)}} & \text{\makecell{ChatGPT \\ Responses}} & 6 & 5 & 4 & 8 & 23\\
\cline{1-7}
\multirow{2}{*}{\textbf{\makecell{Problems \\ found(avg \\per response)}}} & \text{Accuracy} & 0 & 0 & 0.75 & 0.125 & 0.17 \\
&\text{Language} & 0 & 0 & 0.5 & 0.125 & 0.13 \\
& \text{Content} & 1.6 & 0.8 & 2.5 & 1.375 & 1.52 \\
\cline{1-7}
\multirow{2}{*}{\textbf{\makecell{Likert \\ ratings(1-5)}}} & \text{Handle well(n)} & 3.6 & 4.6 & 3.75 & 3.38 & 3.83 \\
& \text{Accept(n)} & 3.16 & 4.4 & 2.75 & 3 & 3.33\\
\bottomrule
\end{tabular}
\caption{Results of annotations for lay people. The statistics represent the count of annotated responses in each scenario. Problems found and the Likert rating are average values. }
\label{patient reported problems}
\end{table*}

\setlength{\tabcolsep}{5pt}
\renewcommand{\arraystretch}{1}
\setlength{\tabcolsep}{2pt}
\begin{table*}[ht]
\centering
\begin{tabular}{ccccccc}
\toprule
&&\textbf{Patient-Explain} & \textbf{Patient-Suggest} & \textbf{Doctor-Explain} &\textbf{Doctor-Suggest} &\textbf{Overall}\\ 
\hline
\multirow{2}{*}{\textbf{Statistics(n)}} & \text{\makecell{ChatGPT \\ Responses}} & 8 & 7 & 5 & 7 & 27\\
\cline{1-7}
\multirow{2}{*}{\textbf{\makecell{Problems \\ found(avg \\per response)}}} & \text{Accuracy} & 0 & 0.86 & 0.6 & 1.29 & 0.67 \\
&\text{Language} & 0.875 & 0.14 & 5 & 1 & 1.48 \\
& \text{Content} & 1.875 & 1.43 & 1.8 & 1.71 & 1.7 \\
\cline{1-7}
\multirow{2}{*}{\textbf{\makecell{Likert \\ ratings(1-5)}}} & \text{Handle well} & 3.5 & 3.29 & 2.4 & 3.43 & 3.16 \\
& \text{Meet standards} & 2.1 & 2.4 & 1.75  & 2.67 & 2.21\\
\bottomrule
\end{tabular}
\caption{ Results of annotations for doctors. The statistics represent the count of annotated responses in each scenario. Problems found and the Likert rating are average values. }
\label{doctor reported problems}
\end{table*}

\textbf{3) Focus groups:}
We ran three focus groups on a workshop to examine and discuss the ChatGPT responses.  The workshop included 23 participants from various backgrounds, including 7 clinicians, 5 computer scientists, 2 researchers, 3 National Health Service (NHS) IT experts, five patients who had experienced cancer and one caregiver (some people had multiple roles and we have only listed their main role). The participants were divided into 3 groups based on their identity, with a similar number of participants from each role.

The participants in the focus groups were shown ChatGPT's responses; they did not see the MDT reports.  Discussions were held separately around benefits and limitations. Every group member illustrated their opinions. The discussion from the focus groups was audio recorded and each group discussion lasted 1 hour. Recordings were transcribed into text and content analysis was used to identify the key themes under discussion; NVivo was used as the tool for coding. 

In addition to the group discussion, we collected individual notes on the following three questions:
(1) Would you ask ChatGPT questions about your own health?
(2) Would you trust the responses?
(3) Do you think ChatGPT would be helpful in cancer care?
Appendix ~\ref {sec:FocusgN} shows the answer provided by participants with role.

\section{Result}
\subsection{Pilot with MDT authors} 
The MDT authors  found several problems with ChatGPT’s explanations of reports, these fell into three main categories:

\textbf{1) Accuracy:} These errors were varied and included:
\begin{itemize}[itemsep=0.5em, topsep=0.5em]
\item {Incorrect interpretation of abbreviations: For example, interpreting "CNP" as "Certified Nurse Practitioner" when it actually referred to a doctor who had these initials.}
\item {Incorrect URL: For example, giving the URL http://www.clanhouse.org for the organisation CLAN cancer support (the actual URL is https://www.clancancersupport.org/)}
\item {Incorrect test results: For example, stating that the current PSA level is 5.0, when the report stated "Your PSA level was 5.0 after surgery it decreased to 0.2"}
\end{itemize}
\textbf{2) Language:} Again, varied errors, including
\begin{itemize}[itemsep=0.3em, topsep=0.3em]
\item {Too complex:  For example, "suprapubic pain” instead of "pain in the lower abdomen".}
\item {Grammar: For example, "its" instead of "it is".}
\item {American English: American spellings (e.g. "organizations") and phrases which are inappropriate in the UK.}
\end{itemize}
\textbf{3) Content:} Content errors included 
\begin{itemize}[itemsep=0.3em, topsep=0.3em]
\item {Too vague: For example, "Connecting with others who are going through similar experiences can provide emotional support and practical advice."}
\item {Not tailored to the patient: For example, ongoing treatments: "Be prepared for a series of treatments, which may include chemotherapy, radiation, or other targeted therapies."}
\item {Too technical for the patient: "The treatment method is neoadjuvant chemoradiotherapy. "}
\end{itemize}

\subsection{Annotation (lay people)}
 Eight lay subjects were asked to annotate problems in ChatGPT responses to prompts, which we categorised into the three categories described above; the subjects collectively annotated 23 ChatGPT responses.  Subjects also rated each response using Likert-scale questions.  Table \ref{patient reported problems} gives statistics of the type of problems they found.  

Overall, 78\% responses had problems, with an average of 1.52 issues in each response.  Responses to Likert questions were overall positive; see Table \ref{patient reported problems}. Subjects thought ChatGPT handled the question well (average Likert score of 3.83/5) but were less willing to accept the answer (average Likert score of 3.33/5) as a patient. One participant strongly disagreed with the answer to the patient-explain questions, and two strongly disagreed with the response to the doctor-suggest questions.

Lay subjects found it difficult to identify inaccurate responses, but did report many content problems, where reports were confusing, used medical jargon, were too generic, and were not aligned with patients' needs.  They also commented on the inappropriate use of the degree adverb (e.g. "Aim to totally cure your cancer"), which could mislead patients about the severity of their condition. New issues and examples can be found in the appendix ~\ref {sec:Layissues}.

\subsection{Annotations (Clinicians)}

Clinicians (including the two MDT authors) also completed the annotation exercise; they analysed twenty-seven ChatGPT responses.  Table \ref{doctor reported problems} summarises what they found.

Overall, doctors found more issues than lay people and gave lower Likert ratings. On average, each response was annotated with 3.85 issues, and a total of 92.59\% of the responses reported problems. Eighteen issues were reported to be inaccurate from twenty-seven responses, leading to risk concerns. Generally, doctors held a positive view that ChatGPT handled the questions well (average Likert score of 3.16), but they believe it's still far away from the medical standard (average Likert score of 2.21).

Problems found included:
\begin{itemize}[itemsep=0.3em, topsep=0.3em]
\item {Accuracy: Wrong medical terms, wrong organisation, wrong dates, misunderstanding indicators, giving suggestions based on the old test results.}
\item {Language: Technical language, showing empathy but being too sentimental for clinical cases, inappropriate expressions that could upset patients (e.g. "It may be helpful to get your affairs in order, including any legal and financial planning, which can provide peace of mind for you and your family"), American terminology (e.g. "External radiation therapy” instead of (UK) "external beam radiotherapy")}
\item {Content: Not tailored to patients (e.g. inappropriate use of "normal"), incorrect or inappropriate suggestions (e.g. "Utilize the multidisciplinary team (MDT) approach"). Details and examples can be found in appendix ~\ref {sec:doctorAiss}.}
\end{itemize}

In general, doctors found the structured summaries that ChatGPT offered to be helpful. Some functions, like helping doctors write letters to patients, could be useful but would need to be heavily screened by the clinician. Both patients and doctors noted that they might start to use the technology informally. However, it is not currently appropriate for clinical implementation due to governance, privacy, security issues, and accuracy.

\subsection{Focus group workshop}
 
Our three focus groups discussed a set of ChatGPT responses (see Appendix  ~\ref {sec:FocusgroupP}) to the same MDT reports. They had some positive comments on ChatGPT, but most of the discussion focused on barriers to adaption, as well as requirements and applications which we summarise below.  Table \ref{tabissue} shows which issues were raised in the different scenarios.

\textbf{Barriers to Patient Adoption}
Patients felt the ChatGPT responses were hard to understand and suggestions were generic when reading them. When discussing applying AI in real health contexts, they were concerned about information security, mistakes and trust. The main barriers identified were:
\begin{itemize}[itemsep=0.3em, topsep=0.3em]
\item {Responses contain lots of medical jargon, which is not easy for lay people to understand.}
\item {ChatGPT is not able to give insights into what the information means to the patient and its implications for that individual.}
\item {Many suggestions were superficial and generic and not helpful to the patient.  }
\item {Information security and privacy concerns. Personal data should not be shared with a third party without protection.}
\item {Almost all cancer patients in the focus groups said they would struggle to trust ChatGPT or to use AI-generated suggestions without cross-referencing the information. }

\textit{Quote: It's when I speak to the doctor. I know that the doctor's been in medical school for seven years and then they've got 20 years experience or whatever, you know, I get confidence. }
\item {Trust was further reduced by obvious mistakes, such as the ones described above.
Patients said that even if the technology improved over time, they would still prefer to search online or ask doctors.}
\end{itemize}

\textbf{Barriers to Clinician Adoption}
There were also many barriers to adoption by clinicians.  They included most of the ones mentioned above and also some additional issues:
\begin{itemize}[itemsep=0.3em, topsep=0.3em]
\item {Integrating ChatGPT into existing clinical workflows, including getting approval from NHS.}
\item {Safety and transparency.}
\item {Responses are not personalized to patients (new/existing patients, educational background, health status, etc).}
\item {Responses use American English spelling and style and are based on the American healthcare system (as mentioned above).}
\end{itemize}

\textbf{Requirements}
The requirements of patients (in addition to addressing the above barriers) were: responses should avoid medical jargon and explain the meaning and impact for that individual.  Also, the system as a whole needs to be user-friendly, especially for elderly cancer patients.

In addition to addressing the barriers mentioned above, doctors expect NLP tools to understand local context and systems, be more patient-friendly and meet the preferences of different users.

\textit{Quote: when you're thinking about writing letters to people, you subconsciously change the tone to suit the patients well, trying and make sure that they understand is very difficult to do.}

Doctors also want the AI system to look at more context and history beyond what is in the MDT report and to avoid over-interpreting the information and being over-empathetic.

Both patients and doctors want doctors to validate responses and suggestions made by the technology.  Some patients remain skeptical about AI and do not wish to be forced to use AI. 

\setlength{\tabcolsep}{5pt}
\renewcommand{\arraystretch}{1}
\begin{table}[ht]
\centering
\begin{tabular}{lccccc}
\toprule
 &\textbf{Issues} &\textbf{Pilot} & \textbf{Lay} & \textbf{Doc} & \textbf{Groups}\\
\hline
1&\makecell[l]{Mistakes} & Y & Y & Y & Y\\
\hline
2&\makecell[l]{Medical jargon} & Y & Y & Y & Y\\
\hline
3&\makecell[l]{American style} & Y &  & Y & Y\\
\hline
4&\makecell[l]{Inappropriate\\empathy} & Y &  & Y & Y\\
\hline
5&\makecell[l]{Too generic} & Y & Y & Y & Y\\
\hline
6&\makecell[l]{Not personalized} & Y &  & Y & Y\\
\hline
7&\makecell[l]{Inappropriate use\\of degree adverbs} &  & Y &  & \\
\hline
8&\makecell[l]{Not align to\\patients need} &  & Y &  & Y\\
\hline
9&\makecell[l]{Need more\\examples} &  & Y & Y & Y\\
\hline
10&\makecell[l]{Expression cause\\ distress for\\cancer patients} &  &  & Y & \\
\hline
11&\makecell[l]{Need more\\clinical meaning\\to the patients} &  &  & Y & Y\\
\hline
12&\makecell[l]{Not aligned\\with clinical \\workflow} &  &  & Y & Y\\
\hline
13&\makecell[l]{Information \\security and \\ privacy} &  &  &  & Y\\
\hline
14&\makecell[l]{Trust} &  & Y & Y & Y\\
\bottomrule
\end{tabular}
\caption{Issues found in different methods}
\label{tabissue}
\end{table}

\section{Discussion}
Cancer patients are increasingly being given access to their own medical records. It is not realistic to expect overworked clinicians to carefully explain and 'translate' complex medical notes to patients.  We hoped that LLM technology could generate more useful responses, that enable patients to get reliable help immediately.

Unfortunately, this is not yet the case. Although ChatGPT often did an impressive job of summarizing information from MDTs in a patient-accessible fashion, many ChatGPT responses were problematic.  Specific issues which repeatedly occurred across our various studies included:
\begin{itemize}[itemsep=0.3em, topsep=0.3em]
\item {Accuracy: Responses were sometimes wrong; errors were medical (e.g., wrong test results) and non-medical (e.g., wrong URL).}
\item {Language: Problems included overuse of medical jargon, Americanisms and over-sentimental language.}
\item {Content: Responses often failed to provide the information that patients wanted to know; sometimes they scared patients unnecessarily.}
\end{itemize}

Trust emerged as a major issue; patients and doctors were reluctant to trust ChatGPT responses unless these had been checked, preferably by clinicians and some patients did not want to use them at all.  Getting approval from the health service to use this technology could also be challenging and would require addressing data privacy and security concerns as well as the quality of accuracy, language and content.

Of course, it is likely that we will get better results if we use techniques such as prompt engineering, fine-tuning and one-shot learning to improve the performance of the language model; we could also add safety-checking tools to detect problems such as spam URLs.    We will pursue this in future work, but it is possible that while such techniques will reduce the frequency of problems, they will not eliminate them.

However, reducing the frequency of problems will make it more realistic to use LLMs to draft patient-facing letters (the Doctor-Explain scenario) which doctors check and edit; we note that both doctors and patients wanted doctors to validate ChatGPT responses.  Checking and editing needs to be realistic for clinicians, which means reducing the number of issues that need to be fixed in each response.

\section{Future work}
Future research should focus on several key areas to enhance the effectiveness and applicability of AI tools like ChatGPT in healthcare: 

Doing more research on what people need:
\begin{itemize}[itemsep=0.3em, topsep=0.3em]
\item {\textit{Address needs of doctors and patients}: Understanding these needs will help tailor AI tools to better serve both groups, ensuring that the AI provides relevant and useful information.} 
\item {\textit{Understanding doctors' approaches without AI}: This helps the application be designed practically. For example, \citep{knoll-etal-2022-user} studied on clinicians note-taking behaviour and how medical note generation software could be used in clinical practice, offers valuable insights. }
\item {\textit{Integrate application to clinical workflow}: Working closely with medical professionals ensures that AI tools align with clinical practices and standards. }
\end{itemize}
Address the issues discussed in the paper:

\begin{itemize}[itemsep=0.3em, topsep=0.3em]
\item {\textit{Information security}: Establishing strong data governance frameworks to safeguard sensitive information and comply with healthcare regulations is crucial. }
\item {\textit{Incorporating more in-context learning}: Understanding the patient’s educational background or previous conversations could help the model personalise communications. }
\item {\textit{Improve technology}: Adapting technologies such as prompt engineering and model fine-tuning to the use case can improve response quality.}
\item {\textit{Human-AI Interaction}: Human doctors can experience treatment fatigue and AI systems can make errors. Finding the right interaction methods is one way to address these issues. }
\end{itemize}

\section{Conclusion}
This paper evaluates ChatGPT’s ability to assist in explaining complex medical reports.  The findings underscore the importance of making AI-generated explanations more accessible and relevant to patients, enhancing trust in AI systems among healthcare providers and patients, and ensuring that such tools can be effectively integrated into clinical practice. We call on NLP researchers and medical professionals to collaborate on studies and developments in this field to create more useful applications.

\section*{Limitations}
Several limitations of our study should be acknowledged. First, recruiting a large number of doctors and patients for annotation posed challenges, limiting the sample size and potentially affecting the generalizability of our findings. Additionally, the characteristics of the focus group participants were not comprehensively collected, which might have introduced bias or limited the representativeness of the results. Last, the AI tool we applied is only the webpage version ChatGPT4, which limits the finding apply to other LLMs. 

\section*{Ethics Statement}
Our experiment obtained two ethical approvals: one from the computer ethics committee to collect annotations from lay people and one from the medical ethics committee.The focus groups were conducted as part of a larger co-design day which was ethically approved (NHS research ethics and NHS R+D). All experiments were conducted with the informed consent of the participants, ensuring full confidentiality of the research data. Although the MDT reports we tested were fabricated, the content was close to real data. Unfortunately, we cannot share these reports, which limits the reproducibility of our experiment.

\bibliography{acl_latex}

\appendix

\section{Appendix}

\subsection{Pilot Prompts examples}
\label{sec:pilot}
\begin{itemize}[itemsep=0.3em, topsep=0.3em]
\item {Hi, I’m 64, and I have been diagnosed with colorectal cancer. I was told I had an MDT meeting about me, and I’m pretty confused about the result. I asked my doctor to give me the MDT report. Feeling a bit down about the lack of information or support.} 
\item {Can my cancer be cured?} 
\item {I had resections before but unfortunately recur, why is that happened and is there something I should be careful in daily life?} 
\item {What are the lymph nodes draining the right colon?} 
\item {How will I die?}
\item {Should I take the chemotherapy?}
\item {What is my prognosis? }
\item {Why do my stroke and diabetes make treatment choices more challenging?}
\item {How complex is my overall health?}
\item {What dose this mean"CT PET 22/2/22- 5 sites of PET Avidity. "}
\item {How should I tell this to my family?}
\item {Hi, I am a doctor and have a Prostate cancer patient report, can you help me manage the information? you will be a very experienced prostate cancer specialist.}
\item {Can you offer a detailed treatment plan with time for this patient, considering she lives in XXX?}
\item {What should I do to release the patient's anxiety and improve his life quality?}
\end{itemize}

\subsection{MDT and Questionnaire example}
\label{sec:MDTQ}
\textbf{Dear colleague,}

Thank you for collaborating with us on our research project. We are interested in understanding how ChatGPT4 can handle complex patient documents. We have used mock reports from cancer multidisciplinary team (MDT) meetings, built to closely resemble real reports, but not using real patient data. We assumed the identity of the patient or doctor and asked ChatGPT a series of questions. We would like your help in identifying:
\begin{itemize}[itemsep=0.3em, topsep=0.3em]

\item {Inaccurate responses}
\item {Problems with the language used in the responses (e.g. too complex or technical, too ‘Americanised’, lack of emotional consideration)}
\item {Other problems with the content of the response (e.g., seems reasonable but did not solve the problems)}
\end{itemize}

We would like you to read the following examples and do three things:

1.	Annotate the ChatGPT output text (not the MDT report) – you might want to print and annotate by hand or use highlighting/underlining or tracked changes and comments in Microsoft Word to highlight specific lines of text where you notice problems. Tell us what the problem is and please give any suggestions for how the ChatGPT response could be improved. (Please identify the problem type as classified above (inaccurate, language, other problems with content)) 

2.	Give the ChatGPT response an overall rating – how well did you think ChatGPT handled the question in real-world cases?

3.	Provide some overall feedback about the response in the comments box, telling us about things you liked or disliked about ChatGPT’s response, adding any details about any problems you found. What is missing?	

\textbf{Case 1: A colorectal cancer MDT report}

\begin{table}[h]
    \centering
    \begin{tabularx}{0.5\textwidth}{X|X}
        \hline
        \textbf{Patients information} & \textbf{...} \\
        \hline
        \textbf{WHO performance Status} & \textbf{...} \\
        \hline
        \textbf{Pertinent clinical history or operation findings and reason for discussion} & \textbf{...} \\
        \hline
        \textbf{Clinical, Radiological and pathology findings} & \textbf{...} \\
        \hline
        \textbf{...} & \textbf{...} \\
        \hline
    \end{tabularx}
    \caption{MDT report structure sample}
    \label{tab:mdt}
\end{table}

\textbf{1.User(patient) prompt:             }                                      

Hi I’m 23 and I have been diagnosed with colorectal cancer. I was told that the hospital team held an MDT meeting about me and I’m pretty confused about the result. I asked my doctor to give me the MDT report. I’m feeling a bit down about lack of information or support.

\textbf{ChatGPT output:                           [PLEASE ANNOTATE BELOW] }

………………

(ChatGPT responses)

\textbf{Overall rating of the response in clinical cases }                                     
\begin{enumerate}
    \item I think ChatGPT handled the question well.
    \begin{itemize}[label=\(\square\), left=2em]
        \item Strongly agree
        \item Agree
        \item Neither agree nor disagree
        \item Disagree
        \item Strongly disagree
    \end{itemize}

    \item I will accept it if my doctor answers like above.
    \begin{itemize}[label=\(\square\), left=2em]
        \item Strongly agree
        \item Agree
        \item Neither agree nor disagree
        \item Disagree
        \item Strongly disagree
    \end{itemize}
\end{enumerate}

\textbf{Overall Feedback and problems}

...

\subsection{Focus groups notes}
\label{sec:FocusgN}
See table \ref{tab:survey}. Note that not all participants provided the feedback.
\setlength{\tabcolsep}{5pt}
\renewcommand{\arraystretch}{1.2} 

\begin{table*}[ht]
    \centering
    \begin{tabularx}{0.7\textwidth}{l|c|c|c|c|c}
        \hline
        \textbf{Identity} & \textbf{Yes} & \textbf{No} & \textbf{With condition} & \textbf{Maybe} & \textbf{Missing} \\
        \hline
        \multicolumn{6}{l}{\textbf{Question 1: Would you ask ChatGPT about your own health?}} \\
        \hline
        Doctors & 2 & & 1 & 1 & \\
        \hline
        Patients & 2 & 2 & & & \\
        \hline
        Computer scientists & & & 1 & & 1 \\
        \hline
        Researcher/NHS IT & & 2 & & & \\
        \hline
        Carer & & & & & 1 \\
        \hline
        \multicolumn{6}{l}{\textbf{Question 2: Would you trust the responses?}} \\
        \hline
        Doctors & & 1 & 3 & & \\
        \hline
        Patients & 1 & 2 & 1 & & \\
        \hline
        Computer scientists & & 1 & 2 & & \\
        \hline
        Researcher/NHS IT & & 1 & 2 & & \\
        \hline
        Carer & & 1 & & & \\
        \hline
        \multicolumn{6}{l}{\textbf{\makecell{Question 3: Do you think technologies like ChatGPT \\ would be helpful in cancer care?}}} \\
        \hline
        Doctors & 3 & & & 1 & \\
        \hline
        Patients & 2 & 2 & & & \\
        \hline
        Computer scientists & 2 & & & & 1 \\
        \hline
        Researcher/NHS IT & 2 & & & 1 & \\
        \hline
        Carer & & & & & 1 \\
        \hline
    \end{tabularx}
    \caption{Focus groups notes statistic results}
    \label{tab:survey}
\end{table*}

\subsection{Lay annotation new issues}
\label{sec:Layissues}
\textbf{Accuracy:} (1) One is giving inferences/suggestions not directly based on the report, and that could be inaccurate.

\textit{e.g. Comments: The report doesn't write cancer stage, but ChatGPT inferred one and reported}

and (2) The other is the use of overly absolute adverbs to describe treatment methods, leading to interpretations as factually incorrect:

\textit{e.g. This approach is aimed at not just treating the cancer but doing so in a way that aims to cure it completely.}

\textbf{Language:} The language problem considered here is inappropriate expression, especially the use of degree adverbs. This may not cause a serious factual error but could mislead patients: 

\textit{e.g. MDT report: PSA 5.0, was 4.5 \& 5.4. ChatGPT: Your PSA levels have been observed around 5.0.}.

\textbf{Content:} A few participants reported that what they wanted to know and what ChatGPT offered did not match. For example, one participant reported he needed a reason why active surveillance was chosen as the best approach rather than presenting "Your medical team is considering all treatment options, including active surveillance". In this case, ChatGPT explained a sentence in the report, but the patient was more interested in knowing the reason and whether he got the correct treatment.

\subsection{Doctor annotation details}
\label{sec:doctorAiss}

\textbf{Accuracy:}  Patient-explain and Doctor-Explain questions have fewer inaccurate issues than Patient-Suggest and Doctor-Suggest questions. (1) The errors in explaining questions mainly indicate the wrong medical term/test in detail/purpose:

\textit{(e.g. The MDT, consisting of specialists across various disciplines including urology, oncology, radiology, and nursing... Doctor: should include pathology).}

Regarding the suggestive questions, there are more issues reported, (2) such as the not accurate resources:

\textit{(e.g. ChatGPT: resources from NHS Scotland doctor: NHS Inform)}

(3) Wrong plan dates, (4) misunderstanding indicators when they update from clinical history, (5) giving suggestions based on the old test results, and (6) misusing overly absolute adverbs to cause a factual error. 

\textbf{Language:} Language is a big concern for the doctors. Not patient-friendly language and the use of the American style for patients in the UK are the most reported problems. (1) Not being patiently friendly involves a few cases, interpreting medical reports using technical language, showing empathy but being too sentimental to use in clinical cases, and using inappropriate expressions that may upset patients. (a) The same as non-medical background annotators reported, many interpreted words are too technical for a patient to understand

\textit{(e.g. The biopsy has confirmed the presence of microacinar adenocarcinoma on the left side of the prostate).}

(b) Showing empathy is a benefit of LLMs but inappropriate in many clinical cases. For example, "Remember you are not alone in this journey" is excessively sentimental and suits companies more than clinicians. "Your courage and strength during these times have been truly admirable." is marked as not necessary to say. (c) One more issue doctors reported is that the expression may upset patients. For instance, ChatGPT: "It may be helpful to get your affairs in order, including any legal and financial planning, which can provide peace of mind for you and your family". A doctor reported it implies the patient is actively dying, and he would imagine that is what patients would read this sentence to mean. This may sound like a friendly expression, but could cause alarm or distress for someone with cancer.
(2) As well as problems with American English style, doctors also noted differences between American and English clinical systems that could lead to problems.

\textit{(e.g. ChatGPT: External radiation therapy. Doctor: should be external beam radiotherapy or just radiotherapy.)}. 

It is important to note that clinical systems vary in different countries, and cultural differences lead to different treatment methods. Speaking the same language doesn't mean communicating with the patients should be the same way. 

\textbf{Content:} Doctors also mentioned that some content is too generic, vague, and not tailored to patients. (1) From the doctor's perspective, the interpretation should involve more about the clinical meaning for the patients considering the patient's information

\textit{(e.g. ChatGPT: Your PSA was 6.2, which is higher than normal. Doctor: I wonder if the normal is taking into consideration the patient's age and treatments/treatment intent. It would be good if the output could tell the person what "normal" is/what it is basing "normal" on. )}. 

(2) When doctors explain medical reports or give suggestions, they will show more examples 

\textit{(e.g. ChatGPT: Our city has various options for mindfulness. Doctor: It would be great to include specific examples or concrete contact details).}

Note that weblinks and contact details are key information. Otherwise, Google would be more helpful.

(3) Wrong suggestions were discovered, and ChatGPT showed a misunderstanding of the clinical process. For instance, ChatGPT suggested the patient do a self-check every day, while doctors reported this would lead to anxiety. Some suggestions seemed reasonable on the surface, but  doctors noted that they would not fit with their current practice.

\textit{(e.g. ChatGPT: I encourage you to schedule an appointment to discuss these findings further and discuss the proposed treatment plan. Doctor: Not sure the GP would be offering this service at this point).}

Also, some suggestions are not relevant to the treatment 

\textit{(e.g. ChatGPT: Utilize the multidisciplinary team (MDT) approach. Doctor: This is not relevant for a GP managing the patient’s anxiety.).}

Another concern is that some suggestions are not evidence-based, which could lead to inaccuracies or untrustworthiness.

\subsection{Focus group prompts}
\label{sec:FocusgroupP}
\textbf{Group 1: Colorectal cancer01}

1.	User (Patient):
Hi, I was diagnosed with colorectal cancer and I have some problems understanding my MDT report. Can you help me explain it? (Assume the report is copied into ChatGPT)

2.	User (Patient):
What does it mean "indicating the tumour was still locally advanced with involvement in nearby lymph nodes but was completely removed (R0 resection)."? Is the tumour removed or not?

3.	User (Doctor):
We live in XXX, what kind of support can I recommend to the patient to alleviate anxiety and improve her life quality?

\textbf{Group 2: Prostate cancer01}

1.	User (Patient):
Hi, I was diagnosed with prostate cancer and I have some problems understanding my MDT report. Can you help me explain it? (Assume the report is copied into ChatGPT)

2.	User (Patient):
I can't understand my management plan, what does it mean "If for XRT, JG to see patient."

3.	User (Doctor):
We live in XXX, this patient lives alone and has some difficulties coming often, what support I can suggest?

\textbf{Group 3: Prostate cancer02}

1.	User (Patient):
Can you explain the Management Plan in my MDT report to me? (Assume the report is copied into ChatGPT)

2.	User (Patient):
Do I need chemotherapy? 

3.	User (Doctor):
Hi, I'm a GP in XXX and I have an MDT report of a patient, can you write a letter to the patient and inform him of the MDT result? Here's the report:

\end{document}